\documentclass[prl,aps,twocolumn,floatfix,showpacs,superscriptaddress]{revtex4}
\usepackage{graphicx}

\begin{document}

\title{Competition between Spin-Orbit Interaction and Zeeman Coupling in Rashba 2DEGs}

\author{F.E. Meijer} \email{f.e.meijer@tnw.tudelft.nl}
\affiliation{Kavli Institute of NanoScience, Delft University of
Technology,  Lorentzweg 1, 2628 CJ Delft, The Netherlands}
\author{A.F. Morpurgo}
\affiliation{Kavli Institute of NanoScience, Delft University of
Technology,  Lorentzweg 1, 2628 CJ Delft, The Netherlands}
\author{T.M. Klapwijk}
\affiliation{Kavli Institute of NanoScience, Delft University of
Technology,  Lorentzweg 1, 2628 CJ Delft, The Netherlands}
\author{T. Koga}
\altaffiliation[present address: ]{Graduate School of Information
Science and Technology, Hokkaido University, Sapporo, Japan.}
\affiliation{NTT Basic Research Laboratories, NTT Corporation,
Atsugi-shi, Kanagawa 243-0198, Japan} \affiliation{PRESTO-Japan
Science and Technology}
\author{J. Nitta}
\affiliation{NTT Basic Research Laboratories, NTT Corporation,
Atsugi-shi, Kanagawa 243-0198, Japan} \affiliation{CREST-Japan
Science and Technology}

\date{\today}

\begin{abstract}
We investigate systematically how the interplay between Rashba
spin-orbit interaction and Zeeman coupling affects the electron
transport and the spin dynamics in InGaAs-based 2D electron gases.
From the quantitative analysis of the magnetoconductance, measured
in the presence of an in-plane magnetic field, we conclude that
this interplay results in a spin-induced breaking of time reversal
symmetry \textit{and} in an enhancement of the spin relaxation
time. Both effects, due to a partial alignment of the electron
spin along the applied magnetic field, are found to be in
excellent agreement with recent theoretical predictions.
\end{abstract}


\maketitle

Achieving control of the orbital motion of electrons by acting on
their spin is a key concept in modern spintronics and is at the
basis of many proposals in the field of quantum
information\cite{Wolf}. Two physical mechanisms are used to
influence the dynamics of the electron spin in normal conductors:
spin-orbit interaction (SOI) and Zeeman coupling. In the presence
of elastic scattering, these two mechanisms affect the spin in
different ways. SOI is responsible for the randomization of the
spin direction whereas the Zeeman coupling tends to align the spin
along the applied magnetic field. Depending on the relative
strength of these interactions, this interplay of SOI and Zeeman
coupling is responsible for the occurrence of a variety of
physical phenomena\cite{phenomena,Zumbuhl-Minkov}.

Quantum wells (QW) that define 2-dimensional electron gases
(2DEGs) are particularly suitable for the experimental
investigation of the competition between SOI and Zeeman coupling,
since they give control over many of the relevant physical
parameters. Specifically, in these systems the SOI strength can be
controlled by an appropriate QW design\cite{Koga} and by applying
a voltage to a gate electrode\cite{Nitta}. The electron mobility
is usually density dependent, so that the elastic scattering time
can also be tuned by acting on the gate. Finally, Zeeman coupling
to the spin can be achieved with minimal coupling to the orbital
motion of the electrons by applying a magnetic field
\textit{parallel} to the conduction plane.

In this Letter we study the competition of SOI and Zeeman coupling
via magnetoconductance measurements in InGaAs-based 2DEGs with
different Rashba SOI strength. From the detailed quantitative
analysis of the weak antilocalization as a function of an applied
in-plane magnetic field ($B_{\parallel}$), we find that the
partial alignment of the spin along $B_{\parallel}$ results in a
spin-induced time reversal symmetry (TRS) breaking, and in an
increase of the spin relaxation time. The increase in spin
relaxation time is found to be quadratic with $B_{\parallel}$, and
strongly dependent on the SOI strength and the elastic scattering
time. For both the spin-induced TRS breaking and the increase in
spin relaxation time we find excellent quantitative agreement with
recent theory. We also show that the quantitative analysis permits
to determine the in-plane g-factor of the electrons.

The three InAlAs/InGaAs/InAlAs quantum wells used in our work are
very similar to those described in detail elsewhere\cite{Koga}.
Here, we recall that each well is designed to have a different
(Rashba) SOI strength. The characteristic spin-split energy
$\Delta$ for the different samples is $\Delta \approx 0.5, 1.5$
and $1.8$meV (in what follows we will refer to these samples as to
samples 1, 2, and 3, respectively). The electron density and
mobility at $V_{gate}=0 V$ are $n \simeq 7 \cdot 10^{15} m^{-2}$
and $\mu \simeq$ 4 $m^2/Vs$. All measurements have been performed
on ($20$ x $80 ~\mu m$) Hall-bar shaped devices, at 1.6K. A 14 T
superconducting magnet is used to generate $B_{\parallel}$ and
home-made split coils mounted on the sample holder are used to
independently control the perpendicular field ($B_{\perp}$). No
significant difference in the results is observed when the
in-plane field is applied parallel or perpendicular to the
direction of current flow.

\begin{figure}[t]
\begin{center}\leavevmode
\includegraphics[width=1\linewidth]{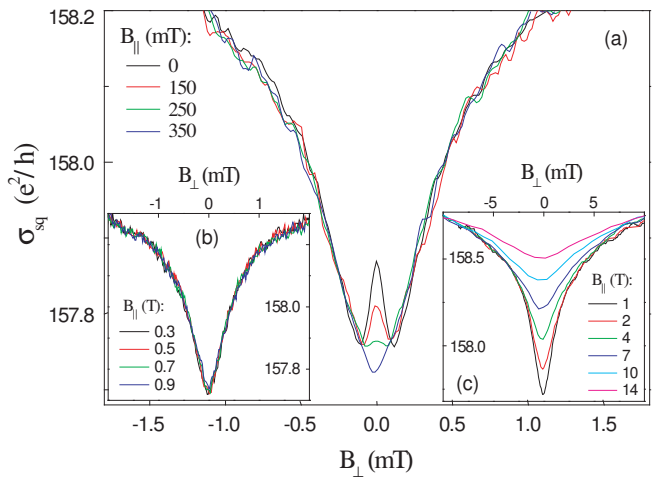}
\caption{The magnetoconductance $\sigma (B_{\perp})$ of sample 1
at different values of $B_{\parallel}$. Three regimes can be
identified: increasing $B_{\parallel}$ from 0 to 350mT results in
a suppression of the WAL peak (a), increasing $B_{\parallel}$
further (up to about $B_{\parallel}=$1T) does not induce
additional changes in the $\sigma (B_{\perp})$-curves (b), for
values of $B_{\parallel}$ larger than 1T the WL is suppressed
(c).} \label{figurename}
\end{center}
\end{figure}

To understand how an in-plane magnetic field affects the
electronic transport, we first discuss the behavior of sample 1
with the weakest SOI strength. Figure 1 shows the
magnetoconductance of this sample measured as a function of
$B_{\perp}$ \cite{x-axis}, for different \textit{fixed} values of
the in-plane field $B_{\parallel}$. For small values of
$B_{\parallel}$ (main panel), the conductance exhibits a maximum
at $B_{\perp}=0$, due to weak-antilocalization (WAL) superimposed
on the background of weak-localization (WL) \cite{WAL}. As
$B_{\parallel}$ is increased, the amplitude of this maximum is
reduced and eventually disappears around $B_{\parallel}= 300$mT. A
further increase in $B_{\parallel}$ does \textit{not} result in
additional changes of the magnetoconductance until $B_{\parallel}$
reaches approximately 1T (fig.1b). Upon increasing $B_{\parallel}$
even further, the WL signal is also suppressed on the scale of
several ($\simeq$ 10) Tesla (fig.1c).

These observations allow us to conclude that the suppression of
WAL and of WL in a parallel field are due to \textit{two distinct
mechanisms} causing time reversal symmetry breaking. At large
fields, $B_{\parallel}\gg 1$T, WL (which is not sensitive to the
spin degree of freedom) is suppressed due to TRS breaking caused
by the coupling of $B_{\parallel}$ to the orbital motion of the
electrons, owing to the finite thickness of the quantum well and
the asymmetric confining potential \cite{TRS}. The suppression of
the WAL peak at smaller values of $B_{\parallel}$ originates from
a spin-induced TRS breaking due to the interplay between
$B_{\parallel}$ (Zeeman coupling) and SOI, as predicted
theoretically\cite{malshukov}. In this paper we will focus on the
spin mechanism for TRS, and discuss the orbital mechanism
elsewhere.

The complete separation of spin and orbital TRS breaking, which is
essential for the work presented here, has not been previously
reported \cite{Zumbuhl-Minkov}. In our samples, this separation is
due to the small QW thickness ($\approx 10nm$) and the small
effective mass ($m^* \approx 0.041 m_0$) which make the subband
splitting in the QW relatively large, as well as to the relatively
large gyromagnetic ratio ($g\simeq 3$)\cite{malshukov,TRS}. It
allows us to account for the magnetoconductance curves
$\sigma(B_{\perp})$ measured at $B_{\parallel} <$ 1T in terms of
existing theories that only consider the coupling of
$B_{\parallel}$ to the electron spin. Therefore, the number of
parameters that need to be introduced for the quantitative
analysis of the data is the smallest possible. This makes it
possible to extract the values of the phase coherence time and the
spin relaxation time as a function of $B_{\parallel}$ with great
accuracy, as it is needed to observe the dependence of $\tau_s$ on
the in-plane magnetic field.

\begin{figure}[t]
\begin{center}\leavevmode
\includegraphics[width=1.0\linewidth]{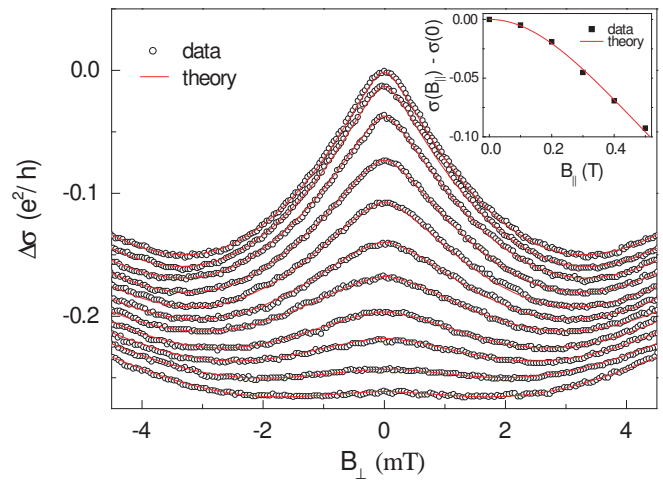}
\caption{The empty circles are the measured magnetoconductance
$\Delta \sigma =\sigma(B_{\perp})-\sigma(0)$ of sample 2 at
different fixed values of $B_{\parallel}$ (offset for clarity).
$B_{\parallel}$ is increased from $0$ to $1$T, in steps of $0.1$T
(top to bottom). The solid lines represent best fits to the ILP
theory. The inset shows the amplitude of the WAL peak at
$B_{\perp}=0$ as function of $B_{\parallel}$, i.e.
$\sigma(B_{\perp}=0, B_{\parallel})-\sigma(0,0)$, and the best fit
to the theory (solid line).} \label{figurename}
\end{center}
\end{figure}

We have performed a quantitative analysis of the
magnetoconductance curves on all samples and for different values
of $n$, by fitting the $\sigma(B_{\perp})$ curves with the theory
of Iordanskii, Lyanda-Geller and Pikus (ILP)
\cite{ILP,Dresselhaus}. This is appropriate for our samples, in
which the spin relaxation is governed by the Dyakonov-Perel
mechanism\cite{DP}. From this analysis, namely from the fits of
$\sigma(B_{\perp})$ curves measured at different values of the
in-plane field, we find the $B_{\parallel}$-dependence  of
$\tau_s$ and $\tau_{\phi}$, i.e. $\tau_s(B_{\parallel})$ and
$\tau_{\phi}(B_{\parallel})$ \cite{phase coherence time}. It is
worth noting that in the ILP theory only one parameter is needed
to account for the spin relaxation, since $\tau_s(0) \equiv
\tau_{s_x} (0)=\tau_{s_y}(0)=2 \tau_{s_z}(0)$. In the presence of
an in-plane field, however, these relations may not hold, since
relaxing the spin along $B_{\parallel}$ costs energy ($\approx g
\mu B_{\parallel}$) whereas relaxation in the direction
perpendicular to $B_{\parallel}$ does not. Nevertheless, for
sufficiently small $B_{\parallel}$ ($g \mu B_{\parallel} < kT$),
the ratios between the different relaxation times are expected to
change only minorly under the conditions of our experiments. This
allows us to treat $\tau_{s} (B_{\parallel})$ as a \textit{single}
fitting parameter.

Figure 2a displays the results of the fitting procedure on sample
2 with the intermediate SOI strength. The continuous lines
superimposed on the data represent the best fit to the ILP theory,
and show that the agreement between data and theory is excellent
for all values of $B_{\parallel}$. Similar agreement is obtained
for the other samples and for all the different values of the
electron density $n$. The values of $\tau_{\phi}(B_{\parallel})$
and $\tau_s(B_{\parallel})$, as extracted from the fits, are shown
in Figs. 3 and 4. Note that, since the electron mobility depends
on the density, we are able to investigate how changing the
elastic scattering $\tau$ affects the $B_{\parallel}$-dependence
of the phase coherence and of the spin-relaxation time. This is of
particular interest as both $\tau_{\phi}(B_{\parallel})$ and
$\tau_s(B_{\parallel})$ are predicted to depend on the
Dyakonov-Perel spin relaxation time $\tau_s(0)$ (see Eqs. 2 and
3), which is related to $\tau$ by the relation
$1/\tau_s(0)=\Delta^2 \tau/2 \hbar^2$ \cite{ILP}.

\begin{figure}[t]
\begin{center}\leavevmode
\includegraphics[width=1\linewidth]{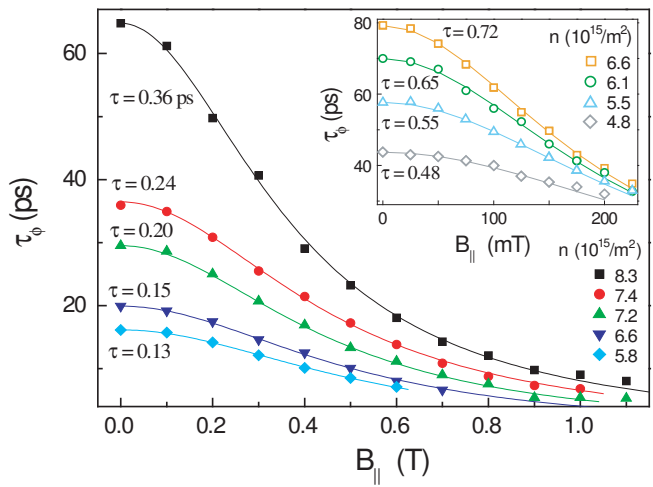}
\caption{The symbols represent $\tau_{\phi}$ as a function of
$B_{\parallel}$, as extracted from the analysis of the
magnetoconductance of sample 2, using the ILP theory (see Fig. 2).
Different curves correspond to different values of $n$ (and
elastic scattering time $\tau$). The solid lines are best fits
based on the theory describing spin-induced dephasing
\cite{malshukov}. The decrease of $\tau_{\phi}$ with decreasing
electron density is consistent with dephasing originating from
electron-electron interaction. The inset shows the extracted
$\tau_{\phi}(B_{\parallel})$ and theoretical fits for sample 1. }
\label{figurename}
\end{center}
\end{figure}

For all values of $n$, the measured $\tau_{\phi}(B_{\parallel})$
decreases as a function of $B_{\parallel}$ (Fig. 3), which shows
quantitatively the breaking of TRS due to the interplay of Zeeman
coupling and SOI. This interplay is predicted to result in a
quadratic dependence of $\tau_{\phi}$ on $B_{\parallel}$
\cite{malshukov}:

\begin{equation} \label{eq:1}
\frac{\tau_{\phi}(B_{\parallel})}{\tau_{\phi}(0)}= \frac{1}{1 + c
B_{\parallel}^2}
\end{equation}

where $c$ is a constant given by:
\begin{equation} \label{eq:2}
c = \tau_{\phi}(0)\tau_{s}(0) (g_{\parallel}^* ~ \mu_B/\hbar)^2
\end{equation}

and $g_{\parallel}^*$ is the in-plane g-factor. The solid lines in
Fig. 3 are best fits to the data using Eq. (1) and treating $c$ as
a (density dependent) fitting parameter. Also in this case the
agreement between experiment and theory is excellent for all
values of $n$ and for the different samples (the inset of Fig. 3
shows the behavior of sample 1. Equally good agreement is found
for sample 3).

\begin{figure}[t]
\begin{center}\leavevmode
\includegraphics[width=1\linewidth]{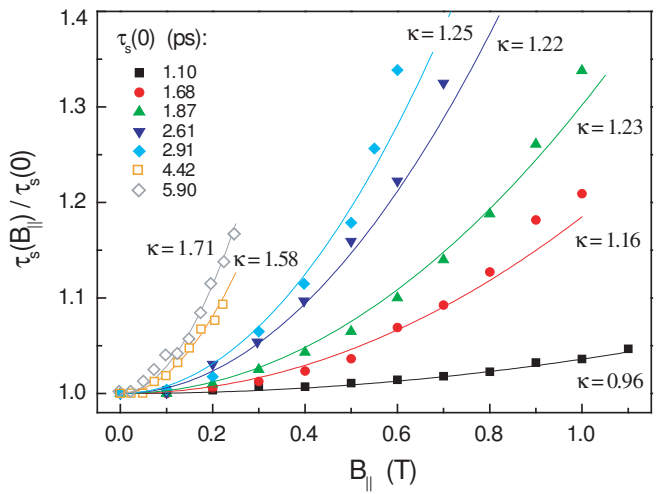}
\caption{The symbols represent $\tau_s$ as function of
$B_{\parallel}$, as extracted from the analysis of the
magnetoconductance of sample 1 (open symbols) and sample 2 (filled
symbols). Each set of symbols corresponds to a different value of
the Dyakonov-Perel spin-relaxation time $\tau_s(0)$, with
$1/\tau_s(0)=\Delta^2 \tau /2 \hbar^2$ (controlled by changing the
gate voltage). The solid lines are best fits to the theory (Eq.
3), with $\kappa$ as an added parameter (see text). Note that the
color and symbol code used in this figure for $\tau_s$ corresponds
to that used in Fig. 3 for $\tau_{\phi}$.} \label{figurename}
\end{center}
\end{figure}

Using the value of $c$ obtained from fitting the data of Fig. 3 we
directly obtain $g_{\parallel}^*$ (Eq. 2). We find that, for each
sample, the in-plane g-factor is approximately constant as a
function of the electron density. The absolute values are
determined to be $|g_{\parallel}^*|=2.8 \pm 0.1$,
$|g_{\parallel}^*|=3.3 \pm 0.1$ and $|g_{\parallel}^*|=3.5 \pm
0.1$, for samples 1, 2 and 3, respectively. Theoretically, the
g-factor in our quantum well is predicted to depend substantially
on its thickness, and is calculated to be $|g_{\parallel}^*|=2.8$
and $|g_{\parallel}^*|=3.5$ for a thickness of 10nm and 15nm,
respectively \cite{Winkler}. This agreement with theory gives
additional support of our analysis in terms of spin-induced
dephasing only, and shows that the measurement of WAL in the
presence of an in-plane field permits to determine the in-plane
g-factor. Contrary to other methods based on transport
measurements, this method to determine the $g$-factor is suitable
for disordered systems.

A different way to obtain $\tau_{\phi}(B_{\parallel})$ (and c),
apart from fitting the whole $\sigma(B_{\perp})$ curves measured
at fixed $B_{\parallel}$, is by looking at the conductance at
$B_{\perp}=0$ as function of $B_{\parallel}$. Specifically, the
theory for spin-induced dephasing predicts that
$\sigma(B_{\perp}=0,0)-\sigma(B_{\perp}=0,B_{\parallel}) =
\frac{e^2}{\pi h} $ln$(\tau_{\phi}(0)/
\tau_{\phi}(B_{\parallel}))=\frac{e^2}{\pi h}$ln$(1 + c
B_{\parallel}^2)$ \cite{malshukov}. Also in this case, the
agreement between theory and data is excellent (Fig. 2, inset) and
the fitting procedure gives values for the parameter $c$ identical
to those obtained above. This shows the consistency of our
quantitative analysis and confirms once more the validity of the
interpretation of the data in terms of spin-induced TRS breaking
only.

Finally, Fig. 4 shows the behavior of the measured spin relaxation
time as function of $B_{\parallel}$ for different densities and
different strength of SOI interaction (samples 1 and 2). In all
cases, the measured spin relaxation time increases quadratically
with increasing the applied in-plane field. This directly shows
that the presence of an in-plane field reduces spin-randomization.
The increase in $\tau_{s}(B_{\parallel})$ is more pronounced for a
small strength of the SOI interaction and for short values of the
elastic scattering time $\tau$, i.e. for long Dyakonov-Perel
spin-relaxation times $\tau_s(0)$. This is because the Zeeman
energy $g \mu B_{\parallel}$, that drives the alignment of the
electron spin along $B_{\parallel}$, competes with the
characteristic energy associated to the spin-randomization
$\hbar/\tau_s(0)$.

A quantitative analysis of the data requires a comparison with
theory. For the case of a magnetic field normal to the conduction
plane, extensive theoretical analysis exists \cite{Burkov}. For
the case of an in-plane field, however, only the relaxation time
of the $z$-component of the spin has been calculated as a function
of $B_{\parallel}$ \cite{Froltsov}. When the Zeeman energy $g \mu
B_{\parallel}$ is much smaller than $\hbar/\tau_s(0)$, this
quantity is given by

\begin{equation}
\frac{\tau_{s_z}(B_{\parallel})}{\tau_{s_z}(0)} \simeq
1+\frac{1}{2} (\kappa g_{\parallel}^* \mu_B B_{\parallel}
\tau_{s}(0)/\hbar)^2
\end{equation}

Although theoretical predictions for $\tau_{s_{x}}(B_{\parallel})$
and $\tau_{s_{y}}(B_{\parallel})$ are not available, we expect
$\tau_{s_{x}}(B_{\parallel})/\tau_{s_{x}}(0)$ and
$\tau_{s_{y}}(B_{\parallel})/\tau_{s_{y}}(0)$ to exhibit the same
functional dependence as
$\tau_{s_{z}}(B_{\parallel})/\tau_{s_{z}}(0)$ as long as $g \mu
B_{\parallel} \ll \hbar/\tau_s(0)$ and $kT$. This allows us to
compare the measured $\tau_s(B_{\parallel})/\tau_s(0)$ to Eq. 3.
All the quantities that appear in Eq. 3 are known from the
previous analysis, and we add a parameter $\kappa$ to achieve best
fits to the data (theory \cite{Froltsov} predicts $\kappa=1$ in
Eq. 3). Figure 4 shows that in all cases good agreement is
obtained with $\kappa \simeq 1$ (continuous lines). We conclude
that the qualitative behavior of the spin-relaxation time as a
function of $B_{\parallel}$, $\tau$ and $\Delta$ (or,
equivalently, $\tau_s(0)$) is the one expected, and that, within a
small correction factor, our results are in quantitative agreement
with theoretical predictions.

In view of the quantitative agreement between theory and data
obtained throughout this work, it is worth considering the origin
of the small correction factor $\kappa$. $\kappa \neq 1$ may
originate from the limited accuracy with which the quantities in
Eq. 3 are determined. The largest uncertainty comes from
$g_{\parallel}^*$ and is approximately 10\%. An additional
possibility is the $B_{\parallel}$-induced anisotropy of the
in-plane spin relaxation times, i.e. $B_{\parallel}$ breaks
spin-rotational symmetry in the 2D plane. Although this anisotropy
is expected to be small for $g_{\parallel}^* \mu B_{\parallel} \ll
\hbar/\tau_s(0)$ and $k_B T$, as mentioned before, it may result
in a deviation from $\kappa = 1$. Finally, for sample 1 with the
weakest Rashba SOI, the Dresselhaus term may not be entirely
negligible \cite{Averkiev}.

In conclusion, we have observed how the partial alignment of the
electron spin along an applied in-plane magnetic field determines
the orbital and spin dynamics of electrons in Rashba 2DEGs. This
alignment results in a spin-induced time reversal symmetry
breaking and in a quadratic increase of the spin-relaxation time.
The detailed quantitative analysis of our results demonstrates the
validity of the existing theory and gives indications to its
limits.

We would like to thank J. Schliemann, B. Nikolic, C.M. Hu, and H.
Takayanagi for stimulating discussion and support. The work of AFM
is part of the NWO Vernieuwingsimpuls 2000 program.


\begin{thebibliography}{9}
\bibitem{x-axis} Experimentally, the value of $B_{\perp}$ is obtained from the measurement of the Hall
voltage performed simultaneously to the measurement of the
longitudinal conductivity. In the presence of a large
$B_{\parallel}$, this automatically takes into account any small
misalignment of the sample.
\bibitem{Dresselhaus} Note that the
Dresselhaus term is much smaller than the Rashba SOI in our
samples (see \cite{Koga}).
\bibitem{phase coherence time} $\tau_{\phi}(B_{\parallel})$ is the upper cut-off time for
interference of time-reversed paths (also denoted as $\tau_{H}$ or
$\tau_{B_{\parallel}}$ in literature (see e.g.
\cite{malshukov,TRS})).
\end{thebibliography}
\end{document}